\documentclass[amsmath,amssymb,pre]{revtex4}

\usepackage[]{graphicx} 
\usepackage{amsmath}

\begin{document}

\title{Transverse density fluctuations around the ground state distribution of counterions near one charged plate: stochastic density functional view}

\author{Hiroshi Frusawa}
\email{frusawa.hiroshi@kochi-tech.ac.jp}

\affiliation{Laboratory of Statistical Physics, Kochi University of Technology, Tosa-Yamada, Kochi 782-8502, Japan.}

\date{\today}

\begin{abstract}
We consider the Dean-Kawasaki (DK) equation of overdamped Brownian particles that forms the basis of the stochastic density functional theory.
Recently, the linearized DK equation has successfully reproduced the full Onsager theory of symmetric electrolyte conductivity. 
In this paper, the linear DK equation is applied to investigate density fluctuations around the ground state distribution of strongly coupled counterions near a charged plate, focusing especially on the transverse dynamics along the plate surface.
Consequently, we find a crossover scale above which the transverse density dynamics appears frozen and below which diffusive behavior of counterions can be observed on the charged plate.
The linear DK equation provides a characteristic length of the dynamical crossover that is similar to the Wigner-Seitz radius used in equilibrium theory for the 2D one-component plasma, which is our main result.
Incidentally, general representations of longitudinal dynamics vertical to the plate further suggest the existence of advective and electrically reverse flows;
these effects remain to be quantitatively investigated.

\vspace{6pt}{\bf Keywords}: stochastic density functional theory; counterions; charged plate; strong coupling; the Wigner-Seitz cell; the Dean-Kawasaki equation
\end{abstract}

%\pacs{}

\maketitle

%%%%%%%%%%%%%%%%%%%%%%%%%%%%%%%%%%%%%%%%%%%%%%%%%%%%%%
%%%%%%%%%%%%%%%%%%%%%%%%%%%%%%%%%%%%%%%%%%%%%%%%%%%%%
\section{Introduction}
Water-soluble materials often have surface chemical groups that are dissociated in a polar solvent.
Examples of such materials include not only mesoscopic particles, such as viruses, proteins, polyelectrolytes, membranes, and micelles, but also macroscopic objects like a glass plate of sample cell [1, 2].
Both mesoscopic and macroscopic particles will be referred to here as "macroions."
The macroions are likely to carry a total surface charge exceeding thousands of elementary charges $e$, surrounded by oppositely charged counterions that are dissociated from the macroions [1, 2];
counterions are electrostatically bound around macroions, due to the high asymmetry between counterions and macroions in valence of charges.

Focusing on the counterions, the macroion systems can be rephrased as inhomogeneous one-component ionic fluids in the presence of external fields:
the one-component fluids of counterions can be regarded as the one-component plasma (OCP) [3-5].
Systems of charged particles immersed in a smooth neutralizing medium are commonly observed in nature, such as a suspension of dust grains in plasmas, as well as colloidal solutions, which can be modeled by the OCP in the unscreened limit of Yukawa fluids [3-5].
The 2D OCP has been used for the description of dusty plasmas confined by external fields, where the motion of particles interacting via 3D electrostatic interaction potential is restricted to a 2D surface  [3-5].
There is, however, a crucial difference between counterion systems and the OCP, due to the localization of the electrically neutralizing background.
While the whole space in the OCP is filled with a smooth background in either the 2D or 3D systems, counterions form a 3D electric double layer and are not neutralized unless they are localized on the macroion surfaces [6-8]. 

This paper will address the strong coupling systems of counterions in the presence of one charged plate, focusing especially on the transverse dynamics of density fluctuations around the ground state that will be specified in Sec. IIA [8-11].
Turning our attention to the dynamics, the strongly coupled counterion system is distinguished from the 2D OCP by one extra dimension, vertical to the macroion surface.
Accordingly, density fluctuations occur not only along the 2D plane parallel to the macroion surface, but also along the one extra dimension. Furthermore, the strong coupling regime in the counterion systems may be realized at room temperature, and therefore dynamics due to thermal fluctuations need to be considered;
however, there are few studies on the counterion dynamics in the ground state.

Thus, the main aim of this paper is to investigate the anisotropic fluctuation field $n({\bf r},t)$ of counterion density due to coarse-grained dynamics of counterion density $\rho({\bf r},t)=\rho_{\infty}({\bf r})+n({\bf r},t)$ around a ground state distribution $\rho_{\infty}({\bf r})$, using the stochastic density functional equation.
The stochastic density functional theory is based on the so-called Dean-Kawasaki (DK) equation that describes the evolution of the instantaneous microscopic density field of overdamped Brownian particles [12-20].
The stochastic density functional theory has been used as one of the most powerful tools for describing slowly fluctuating and/or intermittent phenomena [16-20], such as glassy dynamics, nucleation or pattern formation of colloidal particles; stochastic thermodynamics of colloidal suspensions, dielectric relaxation of Brownian dipoles, and even tumor growth.

The original DK equation includes nonlinear terms of dynamic origin due to the kinetic coefficient that is proportional to fluctuating field $\rho({\bf r},t)$ [12-20].
While the nonlinearility of the original DK equation leads to the above successful descriptions of various phenomena [16-20], a more tractable form is required.
It has been recently demonstrated that the DK equation can be linearized with respect to $n({\bf r},t)$ when $n({\bf r},t)/\rho_{\infty}({\bf r})\ll 1$, and that the linear stochastic equation of density fluctuations is of great practical use [21-27].
The density fluctuations of fluids near equilibrium are surprisingly well described by model-B dynamics of a Gaussian field theory whose effective quadratic Hamiltonian for the density fluctuation field is constructed to yield the exact form of the static density-density correlation function [25].
Furthermore, we have demonstrated that the DK equation can be directly linearized in the first approximation of the driving force due to the free energy functional $F[\rho] $ of an instantaneous density distribution $\rho$, when small density fluctuations around a metastable state are considered \cite{frusawa2019}.

The stochastic thermodynamics around a metastable state has been investigated using the stochastic density functional equation (the DK equation), showing that the heat dissipated into the reservoir is generally negligible \cite{frusawa2019}.
The linear stochastic density functional theory has also been found relevant to investigate out-of-equilibrium phenomena, including the formulations of the full Onsager theory of electrolyte conductivity [22].

The remainder of this paper is organized as follows:
Section II provides formal background in the case of a single charged plate system.
We give the linear DK equation as a stochastic density functional equation, after specifying a general form of the free energy functional $F[\rho]$ of a given density.
In Sec. III, the linear DK equation is applied to the strongly coupled counterion system by considering density fluctuations $n=\rho-\rho_{\infty}$ around $\rho_{\infty}$.
We can verify that the first derivative of $F[\rho]$ in the ground state (i.e., $\delta F[\rho]/\delta\rho|_{\rho=\rho_{\infty}}$) produces a constant, similar to the above metastable state.
Accordingly, the DK equation of the counterion system can be linearized around the ground state.
We will also see the underlying physics of anisotropic fluctuations (vertical to the plate) in terms of the general form of the linear DK equation.
In Sec. IV, we focus on the transverse dynamics along the plate surface, assuming the absence of gradient of fluctuating density field vertical to the plate.
First, we derive the frozen dynamics over a long-range scale beyond the Wigner-Seitz cell, reflecting the formation of Wigner crystal on the charged plate.
Furthermore, the linear DK equation determines a crossover length $l_c$, below which we can observe diffusive behaviors of counterions condensed on the plate.
Our main result in this study is the quantitative evaluation of $l_c$, yielding $l_c\sim a$ for $\Xi\sim10^3$.
Section V contains a summary and conclusions.

%%%%%%%%%%%%%%%%%%%%%%%%%
%%%%%%%%%%%%%%%%%%%%%%%
\section{Formal background}
\subsection{Ground state of counterion system in the strong coupling limit}
Let us briefly summarize what has been achieved by theoretical and simulation studies on the OCP and the counterion systems in the strong coupling regime.

The thermodynamics of the OCP system is characterized by the coupling parameter [3-5],
\begin{equation}
\Gamma=\frac{q^2l_B}{a},
\label{gamma}%%%%%%%%%%%%%%%%%%%%%%%%
\end{equation}
where $q$ is the valence of counterions, $l_B\equiv e^2/4\pi\epsilon k_BT$ is the distance (the so-called Bjerrum length) at which two elementary charges interact electrostatically with thermal energy $k_BT$, when they are surrounded by a polar solvent with its dielectric permittivity and temperature being $\epsilon$ and $T$, and the Wigner-Seitz cell radius $a$ defined by $(\pi a^2)\sigma=q$ using the surface number density $\sigma$ of macroion charges [3-5, 7].
Thermodynamic properties of OCP systems have been extensively studied over decades and accurate numerical results as well as their fits are available in the literature [3-5].
As $\Gamma$ increases, the OCP shows a transition from a weakly coupled gaseous regime ($\Gamma\ll 1$) to a strongly coupled fluid regime ($\Gamma\gg 1$), and it eventually crystallizes.
The concept of the Wigner crystallization due to long-range electrostatic interactions underlies the formation of colloidal crystals, or photonic crystals with large lattice constant, comparable in magnitude to the wavelength of visible light [2-5].

Meanwhile, in counterion systems, a Wigner-Seitz radius $a$ has not been adopted in rescaling the Bjerrum length as $l_B/a$.
We have used another coupling parameter $\Xi$ defined by [6-11]
\begin{align}
&\Xi = \frac{q^2l_B}{\lambda},
\label{xi}\\%%%%%%%%%%%%%%%%%%%%%%
&\lambda=\frac{1}{2\pi ql_B\sigma},
\label{lambda}%%%%%%%%%%%%%%%%%%%%%%%%
\end{align}
using the Gouy-Chapman length $\lambda=1/(2\pi ql_B\sigma)$, a characteristic length of the electric double layer.
Inserting the relation $\sigma=q/(\pi a^2)$ into the definition of $\lambda$, we have
\begin{equation}
\frac{a}{\lambda}=2\pi q\sigma l_Ba=\frac{2q^2l_B}{a}=2\Gamma.
\label{length_comparison}%%%%%%%%%%%%%%%%
\end{equation}
It follows from eqs. (\ref{lambda}) and (\ref{length_comparison}) that
\begin{equation}
\Xi=2\Gamma^2.
\label{gamma xi}
\end{equation}
When macroion-counterion attractions are weak, the structure of such ionic cloud, characterized by $\lambda$, will be dispersed instead of forming the 2D OCP.
The dispersed electric double layer is thus represented by the weak coupling parameter of $\Xi\ll 1$, where the Poisson-Boltzmann approach and its systematic improvements via the loop expansion has been found relevant [6-8]. 
On the other hand, as $\Xi$ increases while reducing $\lambda$, the electric double layer thins and the coarse-grained distribution of counterions becomes two-dimensional [6-11].
Correspondingly, $a$ becomes much larger than $\lambda$ in the strong coupling regime of $\Xi=2\Gamma^2 \gg 1$ (i.e., $a \gg \lambda$), as found from eqs. (\ref{length_comparison}) and (\ref{gamma xi}). 

A field theoretical treatment provides counterion density distribution, $\rho_{\infty}(z_0)$, in the ground state of the strong coupling limit ($\Xi,\,\Gamma\rightarrow\infty$) [6, 8]:
   \begin{eqnarray}
   \rho_{\infty}(z_0)=\frac{\sigma}{\lambda}\exp\,\{-J({\bf r}_0)\,\},
   \label{ground-density}%%%%%%%%%%%%%%%%%%%%%%%%%%
   \end{eqnarray} 
where $J({\bf r}_0)$ denotes the external electrostatic potential in the $k_BT$-unit due to macroion-counterion interactions.
In a single charged plate system, $J({\bf r}_0)$ is expressed as $J({\bf r}_0)=z_0/\lambda$ with $z_0$ denoting the distance between the position ${\bf r}_0$ and the charged plate;
therefore, eq. (\ref{ground-density}) implies that a large portion of the counterions are condensed within the thin electric double layer, which supports the observation that strongly coupled counterions behave like the 2D OCP. 
Extensive Monte Carlo simulations have been performed on the strong coupling regimes of counterions, especially for one-and two-plate systems [8-11].
Accordingly, the asymptotic behavior given by eq. (\ref{ground-density}) has been corroborated by simulation results on the counterion distributions.
The correction to eq. (\ref{ground-density}) has also been evaluated in detail, based on the simulation results.
For about two decades, a variety of strong coupling theories have been developed to explain the above simulation results in the strong coupling regime, focusing not only on the validation of the longitudinal distribution mimicked by the ideal gas behavior (i.e., eq. (\ref{ground-density})) in the vicinity of the charged plate, but also on the deviations from eq. (\ref{ground-density}) for $z_0>\lambda$; see Ref. [8] for a recent review.

%%%%%%%%%%%%%%%%%%%%%%%%%%%
\subsection{Imposing a given density distribution $\rho$ on the grand potential $\Omega$}
Let $\hat{\rho}({\bf r})$ be an instantaneous density of counterions located at ${\bf r}_i\,(i=1,\cdots ,N)$, where the counterion system is rescaled as ${\bf r}=(x,y,z)=(x_0/a,y_0/a,z_0/a)={\bf r}_0/a$.
Figure 1 shows a schematic of the rescaled system.
The instantaneous density is expressed as
\begin{equation}
\hat{\rho}({\bf r})=a^3\sum_{i=1}^N\delta({\bf r}-{\bf r}_i), 
\end{equation}
the use of which counterion-macroion interaction energy for a one-plate system transforms the configurational representation, given by eq. (\ref{delta_ucm}) of Appendix B, to
\begin{align}
\Delta U_{cm}\{\hat{\rho}\}&=\int d{\bf r}J({\bf r})\hat{\rho}({\bf r}),\nonumber\\
J({\bf r})&=2\Gamma z,
\label{ucm-rho}%%%%%%%%%%%%%
\end{align}
which is a functional of $\hat{\rho}$ (see Appendix B for the details).

We can impose a given density distribution $\rho({\bf r})$ on the counterion system via the following delta functional [14, 19-21]:
\begin{equation}
\prod_{{\bf r}}\delta\left[
\hat{\rho}({\bf r})-\rho({\bf r})
\right]
=\int D\psi\,e^{\int d{\bf r}\,\,i\psi({\bf r})\{\hat{\rho}({\bf r})-\rho({\bf r})\}},
\label{rho fourier}
\end{equation}
where the potential field $\psi({\bf r})$ has been introduced in the Fourier transform of the delta functional.
Multiplying the configurational integral representation of the grand potential $\Omega[J]$ (see Appendix B for the definition) by the constraint (\ref{rho fourier}), the formal expression of $F[\rho]$ is obtained:
%%%%%%%%%%%%%%%%%%%%%%(7)
\begin{align}
e^{-F[\rho]}
&=\prod_{{\bf r}}\delta\left[
\hat{\rho}({\bf r})-\rho({\bf r})
\right]e^{-\Omega[J]}\nonumber\\
&=e^{-\int d{\bf r}J({\bf r})\rho({\bf r})}
\int D\psi\,e^{-\Omega[-i\psi]-\int d{\bf r}\,\,i\psi({\bf r})\rho({\bf r})},
\label{appendix constraint}
\end{align}
where $J({\bf r})$ given in eq. (\ref{ucm-rho}) represents the external potential created by the charged plate.
In the mean-field approximation, we obtain (see also Appendix C)
\begin{align}
F[\rho]&=\mathcal{A}[\rho]+\int d{\bf r}J({\bf r})\rho({\bf r}),\label{approximate f rho}\\
\mathcal{A}[\rho]&=\Omega[\psi^*]-\int d{\bf r}\psi^*({\bf r})\rho({\bf r}),\label{hk free energy}
\end{align}
where the saddle-point potential field $\psi^*$ satisfies the following relation:
%%%%%%%%%%%%%%%%%%%%%%(2)
\begin{equation}
\left.
\frac{\delta\left(\beta\Omega[-i\psi]\right)}{\delta\psi({\bf r})}\right|_{\psi=i\psi^*}
=-i\rho({\bf r}).
\label{sp}
\end{equation}
The first Legendre transform of $\Omega[\psi^*]$ provides the Hohenberg-Kohn free energy $\mathcal{A}[\rho]$ defined by eq. (\ref{hk free energy}), the central functional of the equilibrium density functional theory \cite{evans, hansen}.

%%%%%%%%% FIG1 %%%%%%%%%%%%%%
\begin{figure}[htp]
\begin{center}
	\includegraphics[
	width=9 cm
]{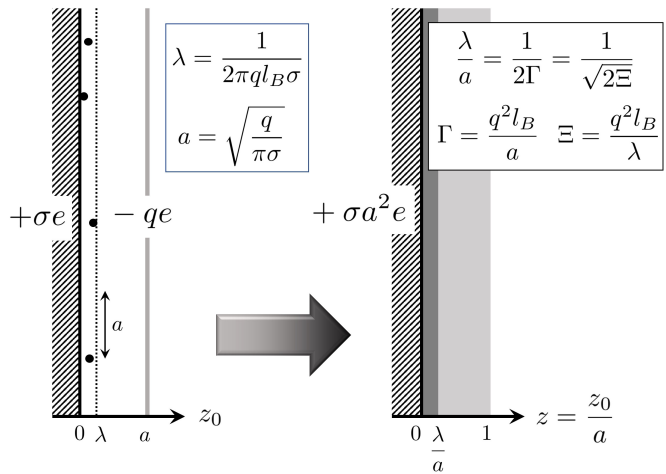}
\end{center}
	\caption{Two schematics illustrating the side view of one charged plate system that consists of a positively charged plate carrying a surface charge density $+\sigma e$ and negatively charged counterions with valence $q$. Our scaling ($z=z_0/a$) of an actual system depicted on the left side implies a coarse-grained system on the right hand side, where $a$ denotes a mean separation between counterions, provided that all of the counterions are condensed on the oppositely charged plate uniformly: $\pi a^2\sigma=q$. The Gouy-Chapman length $\lambda\equiv 1/(2\pi ql_B\sigma)$ is also indicated. This paper adopts the coupling constant $\Gamma$, defined by $\Gamma=q^2l_B/a$ that applies to the 2D OCP, instead of the conventional one, $\Xi=q^2l_B/\lambda$, used for the counterion system.
}
\end{figure}
%%%%%%%%%%%%%%%%%%%%%
%%%%%%%%%%%%%%%%%%%%%%%%%%%%%
%%%%%%%%%%%%%%%%%%%%%%%%%%%%%
\subsection{Stochastic density dynamics obeying the Dean-Kawasaki equation}
Here we focus on the stochastic dynamics of a density field at time $t$, $\rho({\bf r},t)$, whose spatially varying distribution is the same as the coarse-grained variation of $\hat{\rho}({\bf r})$.
What matters in terms of the stochastic density dynamics is the free energy functional $F[\rho]$ of a given density field $\rho$, rather than the grand potential $\Omega$ in equilibrium.
For the density functional, we have provided the approximate form of $F[\rho]$.
The driving force due to $F[\rho]$ and the density-dependent multiplicative noise $\zeta[\rho,\vec{\eta}]$ creates the stochastic dynamics that obeys the DK equation [13-20]:
\begin{equation}
\partial_t\rho=\nabla\cdot D\rho\,\nabla\frac{\delta F[\rho]}{\delta\rho}+\zeta[\rho,\vec{\eta}],
\label{sdft0}%%%%%%%%%%%%%
\end{equation}
where we have introduced a scaled diffusion constant, $D=D_0/a^2$, using the bare diffusion constant $D_0$ and the spatio-temporal average of the multiplicative noise correlations is given by
\begin{flalign}
&\left<
\,\zeta[\rho({\bf r},t),\vec{\eta}({\bf r},t)]\>
\zeta[\rho({\bf r}',t'),\vec{\eta}({\bf r}',t')]
\,
\right>
=-2D\delta(t-t')\nabla_{{\bf r}}\cdot\rho({\bf r},t)\nabla_{{\bf r}}\delta({\bf r}-{\bf r}'),
\label{noise_correlation}%%%%%%%%%%%%%
\end{flalign}
with the vectorial white noise field $\vec{\eta}({\bf r},t)$ that has the correlation $\left<\eta_l({\bf r},t)\eta_m({\bf r}',t')\right>=\delta_{lm}\delta({\bf r}-{\bf r}')\delta(t-t')$.
Equation (\ref{noise_correlation}) can read [13-27]
\begin{equation}
\zeta[\rho,{\bf \eta}]=\nabla\cdot\sqrt{2D\rho({\bf r},t)}\vec{\eta}({\bf r},t).
\end{equation}
In general, the stochastic equation (\ref{sdft0}) includes not only the multiplicative noise term, but also the nonlinear term associated with $F[\rho]$.
As shown below, however, a linear DK equation may be used to investigate the stochastic density dynamics due to fluctuations of counterions in the ground state by expanding eq. (\ref{sdft0}) around $\rho_{\infty}$.

%%%%%%%%%%%%%%%%%%%%%%%%%%%%%%%%%%%%%%%%%%%%%%%%%%%%%%%%%%%%%%%%%%%%%%%%%%%%%%%%%%%%
\section{Stochastic density functional equation for fluctuations around the ground state distribution $\rho_{\infty}$}
\subsection{Linearizing the stochastic Dean-Kawasaki equation (\ref{sdft0})}
Expanding the first derivative of $F[\rho]$ around $\rho_{\infty}$, we have
\begin{flalign}
\nabla\frac{\delta F[\rho]}{\delta\rho({\bf r},t)}
&=\nabla\left[\left.
\frac{\delta F[\rho]}{\delta\rho({\bf r},t)}\right|_{\rho=\rho_{\infty}}
+\int d{\bf r}'\left.
\frac{\delta^2 F[\rho]}{\delta\rho({\bf r},t)\delta\rho({\bf r}',t)}\right|_{\rho=\rho_{\infty}}
n({\bf r}')\right]\nonumber\\
&=\nabla\left[-\int d{\bf r}'
c({\bf r}-{\bf r}')n({\bf r}',t)+\frac{n({\bf r},t)}{\rho_{\infty}(z)}\right],
\label{1-derivative-dynamics}\\%%%%%%%%%%%%%
\rho({\bf r},t)&=\rho_{\infty}({\bf r})+n({\bf r},t),
\label{fluctuation}
\end{flalign}
due to
\begin{equation}
\nabla\cdot
D\rho({\bf r})\left.\nabla\frac{F[\rho]}{\delta\rho}\right|_{\rho=\rho_{\infty}}=0;
\label{constant first derivative}
\end{equation}
see Appendix D for the details.
It follows from eqs. (\ref{1-derivative-dynamics}) and (\ref{constant first derivative}) that the right hand side (rhs) of eq. (\ref{sdft0}) reads
\begin{flalign}
\nabla\cdot D\rho\nabla\frac{\delta F[\rho]}{\delta\rho({\bf r},t)}+\zeta[\rho,\vec{\eta}]
&=\nabla\cdot D\rho_{\infty}\left(
1+\frac{n({\bf r},t)}{\rho_{\infty}}\right)\nabla\left\{
-\int d{\bf r}'
c({\bf r}-{\bf r}')n({\bf r}',t)+\frac{n({\bf r},t)}{\rho_{\infty}(z)}
\right\}
+\zeta\left[\rho_{\infty}\left(
1+\frac{n({\bf r},t)}{\rho_{\infty}}\right),\vec{\eta}
\right]\nonumber\\
&\approx\nabla\cdot D\rho_{\infty}\nabla\left\{
-\int d{\bf r}'
c({\bf r}-{\bf r}')n({\bf r}',t)+\frac{n({\bf r},t)}{\rho_{\infty}(z)}
\right\}+\zeta[\rho_{\infty},\vec{\eta}],
\label{rhs-DK}
\end{flalign}
when $n/\rho_{\infty}\ll 1$.
It is to be noted that the ideal gas distribution $\rho_{\infty}({\bf r})$ given by eq. (\ref{ground-density}) reproduces only the longitudinal distribution of simulation results in the vicinity of a highly charged plate, resulting from attractive and repulsive Coulomb interactions in the strong coupling limit (see also Appendix A).
Correspondingly, eq. (\ref{rhs-DK}) reveals that the dynamics of fluctuating density $n({\bf r},t)$ is governed by the strong Coulomb interactions as represented by the contribution, $-\int d{\bf r}'c({\bf r}-{\bf r}')n({\bf r}',t)$, on rhs of eq. (\ref{rhs-DK}).
In eqs. (\ref{1-derivative-dynamics}) and (\ref{rhs-DK}), the direct correlation function $c({\bf r}-{\bf r}')$ appears because $F[\rho]$ is expressed using the Hohenberg-Kohn free energy functional $\mathcal{A}[\rho]$ as given by eq. (\ref{hk free energy}) \cite{evans, hansen}.
We also have
\begin{equation}
\partial_t\{\rho_{\infty}(z)+n({\bf r},t)\}
=\partial_tn({\bf r},t),
\label{left-DK}
\end{equation}
due to $\partial_t\rho_{\infty}=0$, on the left hand side of the DK equation (\ref{sdft0}).

Combining eqs. (\ref{sdft0}), (\ref{rhs-DK}) and (\ref{left-DK}), we obtain the linear DK equation:
\begin{flalign}
\partial_t n({\bf r},t)&=D\nabla^2n({\bf r},t)
+D\rho_{\infty}(z)\nabla^2\psi_n({\bf r},t)
-\nabla\cdot{\bf j}_{\perp}({\bf r},t)+\zeta[\rho_{\infty},\vec{\eta}],
\label{linearized-DK}
\end{flalign}
where $\psi_n$ denotes a fluctuating Coulomb potential defined by
\begin{equation}
\psi_n({\bf r},t)\equiv-\int d{\bf r}'
c({\bf r}-{\bf r}')n({\bf r}'),
\label{psi_n}
\end{equation}
and the longitudinal current ${\bf j}_{\perp}({\bf r},t)=(0,\,0,\,j_z)$, which is along the $z$-axis vertical to the charged plate, arises from the gradient of the ground density distribution $\nabla\rho_{\infty}$.
Incidentally, in the rescaled system of $\rho_{\infty}({\bf r},t)=a^3\rho_{\infty}({\bf r}_0,t)$, eq. (\ref{ground-density}) is rewritten as
\begin{equation}
\rho_{\infty}(z)=\frac{\sigma a^3}{\lambda}e^{-2\Gamma z},
\label{rescaled ground density}
\end{equation}
thereby providing
\begin{equation}
\nabla\rho_{\infty}=(0,\,0,\,-2\Gamma\rho_{\infty})
\label{gradient}
\end{equation}
in the rescaled system.
In the next subsection, we will discuss the details of ${\bf j}_{\perp}({\bf r},t)$ associated with the presence of $\nabla\rho_{\infty}$ in the longitudinal direction.

Equation (\ref{linearized-DK}) appears to be a simple extension of the Poisson-Nernst-Planck equation [30] when the longitudinal current ${\bf j}_{\perp}({\bf r},t)$ disappears.
However, in actuality, the Poisson-like equation in the second term on the rhs of eq. (\ref{linearized-DK}) differs from the conventional Poisson equation because the interaction potential is replaced by the direct correlation function. 
In this paper, we adopt the direct correlation function, being of the following form [31-33]:
%%% DCF
\begin{flalign}
-c({\bf r})&=\int d{\bf r}'\frac{\Gamma}{|{\bf r}-{\bf r}'|}g_a({\bf r}')=\Gamma\,\widetilde{v}_L(r),
\label{erf potential}\\%%%%%%%%%%%%%%%%%%%%
g_a({\bf r}-{\bf r}')&=\frac{e^{-|{\bf r}-{\bf r}'|^2/m}}{(m\pi)^{3/2}}=\frac{e^{-|{\bf r}_0-{\bf r}_0'|^2/(ma^2)}}{(ma^2\pi)^{3/2}}\>(m=1.08^{-2}),
\label{gaussian}\\%%%%%%%%%%%%
\widetilde{v}_L(r)&=\frac{\mathrm{erf}(1.08 r)}{r}\,(r\equiv|{\bf r}|=|{\bf r}_0|/a),
\label{vl}
\end{flalign}
which gives
\begin{flalign}
\nabla^2\psi_n({\bf r},t)&\equiv
-\int d{\bf r}'
\nabla^2c({\bf r}-{\bf r}')n({\bf r}',t)\nonumber\\
&=-4\pi\Gamma
\int d{\bf r}' g_a({\bf r}-{\bf r}')n({\bf r}',t)\nonumber\\
&=-4\pi\Gamma\widetilde{n}({\bf r},t),
\label{poisson like}%%%%%%%%%%%%
\end{flalign}
where $\widetilde{n}({\bf r},t)$ denotes a coarse-grained density that is smeared by the Gaussian distribution function $g_a({\bf r}-{\bf r}')$ over a range of the Wigner-Seitz radius $a$.
The above form of the direct correlation function has been demonstrated to be available for the OCP in the strong coupling regime of $\Gamma\gg 1$.
In eq. (\ref{erf potential}), the bare electrostatic potential ($\sim 1/r$) is modified using the Gaussian distribution function $g_a({\bf r})$, and the second equation of eq. (\ref{erf potential}) introduces the function of $\widetilde{v}_L(r)=\mathrm{erf}(1.08 r)/r$ that represents the long-range part of the Coulomb interaction potential [31, 32].
It is to be noted that the internal energy of the OCP, obtained using this direct correlation function, or eq. (\ref{erf potential}), exhibits an error of less than $0.8\%$ in the strong coupling regime [31].

%%%%%%%%%%%%
Considering that the Fourier transform $\widetilde{v}_L(k)$ of $\widetilde{v}_L(r_0/a)$ is given by
\begin{equation}
\widetilde{v}_L(k)=\frac{4\pi}{k^2}e^{-k^2(ma^2)/4},
\label{fourier}
\end{equation}
eq. (\ref{poisson like}) is rewritten in the original coordinate of ${\bf r}_0$ as
\begin{align}
\nabla^2\psi_n({\bf r}_0,t)&=-k^2\psi_n(k,t)\nonumber\\
&=-4\pi q^2l_Be^{-k^2(ma^2)/4}\,n(k,t),
\label{poisson fourier}
\end{align}
using the Fourier transforms of $\psi_n({\bf r}_0,t)$ and $n({\bf r}_0,t)$: $\psi_n(k,t)$ and $n(k,t)$.
In the limit of $a\rightarrow 0$, eq. (\ref{poisson like}) is reduced to the conventional Poisson equation.
It follows from eqs. (\ref{poisson like}) and (\ref{poisson fourier}) that the Fourier transform $\widetilde{n}(k,t)$ of the coarse-grained density $\widetilde{n}({\bf r}_0,t)$ reads
\begin{align}
\widetilde{n}(k,t)=e^{-k^2(ma^2)/4}\,n(k,t),
\label{coarse grain}
\end{align}
implying the cut-off of $\widetilde{n}(k,t)$ at a high wavenumber in correspondence with the coarse-graining of $n({\bf r},t)$.

%%%%%%%%%%%%%%%%%%%%%%%%%%%%%
%%%%%%%%%%%%%%%%%%%%%%%%%%%%%
\subsection{Implications of longitudinal contributions given by eq. (\ref{vertical})}
As described in Sec. IIIA, the gradient of $\rho_{\infty}({\bf r})$ has only the $z$-component as found from eq. (\ref{gradient}).
The $z$-component given by $\partial_z\rho_{\infty}=-2\Gamma\rho_{\infty}$ yields the longitudinal contribution, $-\nabla\cdot{\bf j}_{\perp}({\bf r},t)$, to the rhs of eq. (\ref{linearized-DK}):
\begin{align}
-\nabla\cdot{\bf j}_{\perp}({\bf r},t)&=-\partial_zj_z({\bf r},t)\nonumber\\
&=2\Gamma D\partial_zn({\bf r},t)-D\partial_z\rho_{\infty}(z)\Gamma\mathcal{E}_z({\bf r},t)\nonumber\\
&=2\Gamma D\left\{
\partial_zn({\bf r},t)+\rho_{\infty}(z)\Gamma\mathcal{E}_z({\bf r},t)
\right\},
\label{vertical}
\end{align}
where $\Gamma\mathcal{E}_z({\bf r},t)$ denotes the $z$-component of fluctuating electric field ${\bf E}({\bf r},t)$ defined by
\begin{equation}
{\bf E}({\bf r},t)=-\nabla\psi_n({\bf r},t)
=\Gamma\left[\begin{array}{c}\mathcal{E}_x({\bf r},t)\\\mathcal{E}_y({\bf r},t)\\\mathcal{E}_z({\bf r},t)\end{array}\right].
\label{e-field}%%%%%%%%%%%%%%%%%
\end{equation}
It is found from eqs. (\ref{psi_n}), (\ref{vertical}) and (\ref{e-field}) that $-\nabla\cdot{\bf j}_{\perp}({\bf r},t)$ disappears in the absence of the longitudinal variance in $n({\bf r},t)$ (i.e., $\partial_zn({\bf r},t)=0$) as well as $\rho_{\infty}({\bf r})$, which is the reason why ${\bf j}_{\perp}({\bf r},t)$ has been referred to as the longitudinal current.
The two terms on the rhs of eq. (\ref{vertical}) are expressed in the original coordinate as follows:
\begin{flalign}
&2\Gamma D\partial_z n({\bf r},t)=\frac{a}{\lambda}\left(\frac{D_0}{a^2}\right)
\partial_z n({\bf r},t)=\frac{D_0}{\lambda}\partial_{z_0} n({\bf r}_0,t),
\label{long1}\\
&2\Gamma D\rho_{\infty}(z)\Gamma\mathcal{E}_z({\bf r},t)
=\frac{a}{\lambda}\left(\frac{D_0}{a^2}\right)\rho_{\infty}(z)q^2l_B\mathcal{E}_{z}({\bf r}_0,t)
=D_0\rho_{\infty}(z)\left\{\frac{q^2l_B\mathcal{E}_{z_0}({\bf r}_0,t)}{\lambda}\right\}.
\label{long2}
\end{flalign}
Equation (\ref{long1}) represents a vertical advection of fluctuating density field $n({\bf r}_0,t)$.
Putting this advection term given by eq. (\ref{long1}) on the left hand side of eq. (\ref{linearized-DK}), combination of eqs. (\ref{vertical})--(\ref{long2}) transforms  eq. (\ref{linearized-DK}) to
\begin{flalign}
\partial_t n({\bf r}_0,t)\underline{-\frac{D_0}{\lambda}\partial_{z_0} n({\bf r}_0,t)}
=D_0\nabla^2n({\bf r}_0,t)-D_0q^2l_B\rho_{\infty}({\bf r}_0,t)\left\{
4\pi\widetilde{n}({\bf r}_0,t)\underline{-\frac{\mathcal{E}_{z_0}({\bf r}_0,t)}{\lambda}}
\right\}+\zeta[\rho_{\infty},\vec{\eta}]
\label{original stochastic}
\end{flalign}
in the original coordinate representation, where the underlined terms corresponding to the longitudinal contributions.
The former contribution, the second term on the left hand side of eq. (\ref{original stochastic}), suppresses density fluctuations, whereas the latter, the third term on the rhs of eq. (\ref{original stochastic}), acts as a positive feedback to enhance counterion condensation in proximity to the charged plate.
Figure 2 is a schematic of such opposite roles of longitudinal contributions from the above underlined terms.

On the one hand, the underlined term on the left hand side of eq. (\ref{original stochastic}) represents the advective flow term.
The advection velocity is given by $D_0/\lambda$, which increases as the Gouy-Chapman length $\lambda$, a characteristic length of the electric double layer, is shorter. 
The negative sign of this term indicates that the flow direction is always in the opposite direction to the $z$-axis.
Figure 2 illustrates translation of whole fluctuating density field $n({\bf r},t)$, like a Goldstone-mode, due to the advection flow.
Figure 2 shows the case of $\partial_{z_0}n<0$ where the increase from $\rho_{\infty}(0)$ (i.e., $n({\bf r}_0,t)>0$ at $z_0=0$) is lowered when $\partial_{z_0}n<0$, and vice versa because of the fixed direction of the advection flow. 
In other words, density fluctuations are suppressed due to the former longitudinal contribution associated with advection flow.

On the other hand, the underlined contribution of the third term on the rhs of eq. (\ref{original stochastic}) arises from the $z_0$-component $\Gamma\mathcal{E}_{z_0}$ of a fluctuating electric field ${\bf E}({\bf r}_0,t)$;
however, this term reduces the third term on the rhs of eq. (\ref{original stochastic}), or the smeared density $\widetilde{n}({\bf r}_0,t)$ induced by ${\bf E}({\bf r}_0,t)$ itself.
We should remember that, in the limit of $a\rightarrow 0$, the second term on the rhs of eq. (\ref{original stochastic}) corresponds to the electrostatic term that is associated with the Poisson equation:
we have $-D_0q^2l_B\rho_{\infty}({\bf r}_0,t)4\pi n({\bf r}_0,t)$ with $\widetilde{n}({\bf r}_0,t)$ replaced by the original density $n({\bf r}_0,t)$, indicating the electrostatically restoring term to $n({\bf r}_0,t)\rightarrow 0$ as represented by the negative sign.
Accordingly, the latter longitudinal contribution plays a role of positive feedback for counterion condensation, as opposed to the above electrostatic suppression.
In actuality, the latter term increases the strength of longitudinal fluctuating field when $\mathcal{E}_{z_0}>0$, or $\partial_{z_0}n({\bf r}_0,t)<0$, which means that counterions have become more condensed.
Meanwhile, the negative sign of $\mathcal{E}_{z_0}<0$ yields the restoring contribution to $n({\bf r}_0,t)\rightarrow 0$ when accumulated counterions leave the charged plate: $\partial_{z_0}n({\bf r}_0,t)>0$.
In both cases of $\mathcal{E}_{z_0}>0$ and $\mathcal{E}_{z_0}<0$, the latter longitudinal enhances counterion condensation, which is represented as electrically reverse flow for $\partial_{z_0}n({\bf r}_0,t)<0$ in Fig. 2.

%%%%%%%%% FIG2 %%%%%%%%%%%%%%
\begin{figure}[hbtp]
\begin{center}
	\includegraphics[
	width=11.5 cm
]{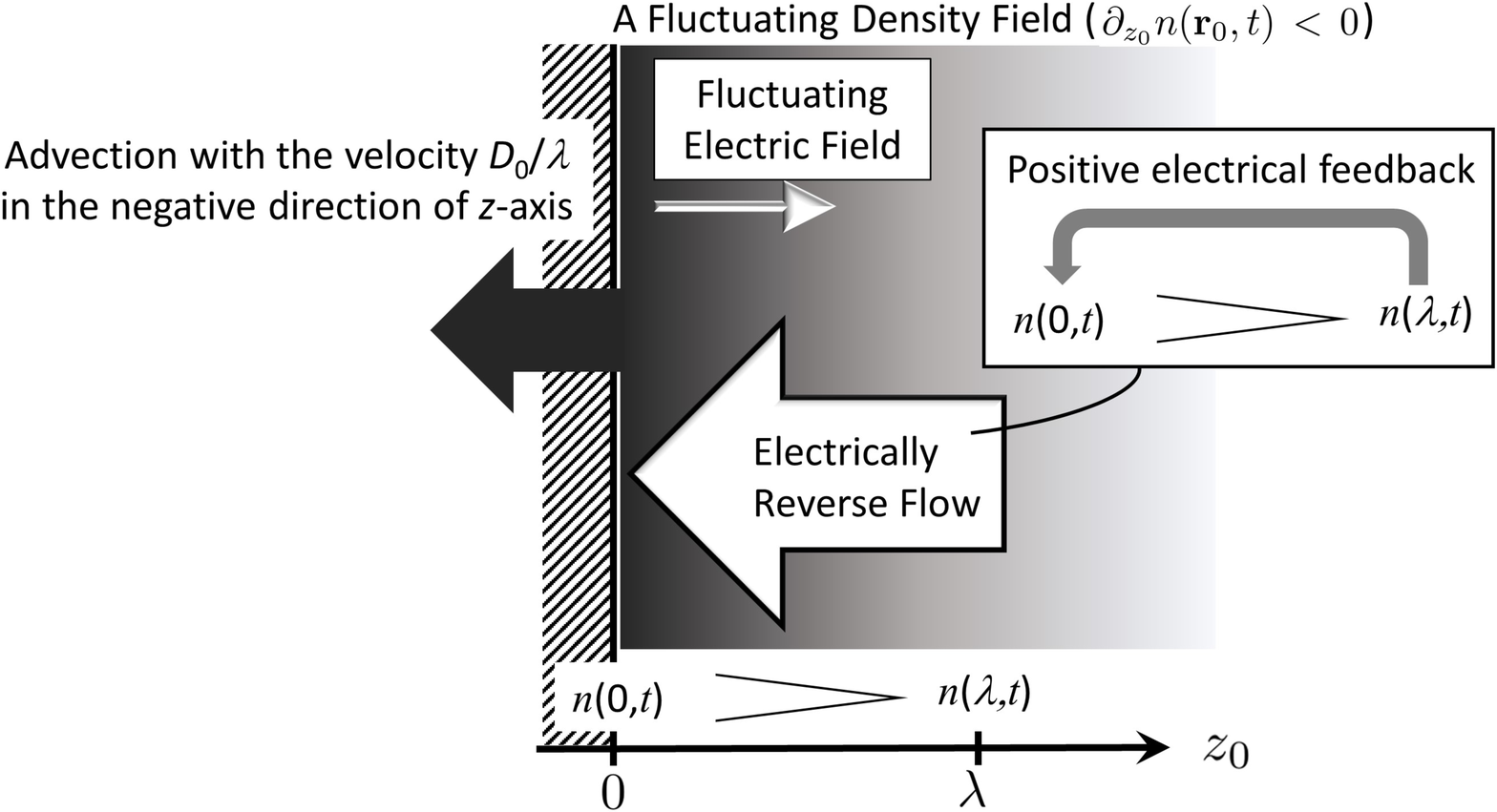}
\end{center}
\caption{
A schematic of anisotropic fluctuations due to longitudinal terms underlined in eq. (\ref{original stochastic}).
Here we consider the case that the fluctuating density $n({\bf r}_0,t)$ decreases with $z_0$, and $n(0.t)$ and $n(\lambda,t)$ are abbreviations of $n({\bf r}_0,t)|_{z_0=0}$ and $n({\bf r}_0,t)|_{z_0=\lambda}$, respectively.
The advective flow, or migration of density fluctuations as a whole, is always in the negative direction along the $z_0$-axis.
Meanwhile, the electrically reverse flow is also in the negative direction, irrespective of the sign of $\partial_{z_0}n$, or $\mathcal{E}_z$.
Accordingly, the latter flow acts as a positive feedback of density fluctuations for $\partial_{z_0}n<0$.
}
\end{figure}
%%%%%%%%%%%%%%%%%%%%%

%%%%%%%%%%%%%%%%%%%%%%%%%%%%%
%%%%%%%%%%%%%%%%%%%%%%%%%%%%%
\section{Density-density correlations due to transverse dynamics along the plate surface}
Supposing that $\partial_{z_0}n({\bf r}_0,t)=0$, we can focus on the transverse dynamics parallel to the charged plate at $z_0=0$, which we will investigate quantitatively.
With the use of the Fourier transform $n(k,t)$ of $n({\bf r}_0,t)$, eq. (\ref{original stochastic}) becomes
\begin{flalign}
\partial_t n(k,t)
&=-D_0\left\{k^2+4\pi q^2l_B\rho_{\infty}(0) e^{-k^2(ma^2)/4}
\right\}n(k,t)+\zeta[\rho_{\infty},\vec{\eta}]\nonumber\\
&=-D_0\Xi\,\mathcal{G}(k)n(k,t)+\zeta[\rho_{\infty},\vec{\eta}],\label{fourier transverse}\\
\mathcal{G}(k)&=\frac{k^2}{\Xi}+4\pi\sigma e^{-k^2(ma^2)/4},
\label{propagator}
\end{flalign}
given that $\partial_{z_0}n({\bf r}_0,t)=0$.
In eq. (\ref{propagator}), the conventional coupling constant $\Xi=q^2l_B/\lambda$ given by eq. (\ref{xi}) appears using $\rho_{\infty}(0)=\sigma/\lambda$.
In both of the strong coupling regime ($\Xi\gg 1$) and the low wavenumber region ($ka\ll 1$), the propagator $\mathcal{G}(k)$ is approximated by
\begin{flalign}
\mathcal{G}(k)\approx\frac{1}{a^2}\left\{
\frac{(ka)^2}{\Xi}+4q
\right\}\approx \frac{4q}{a^2}=4\pi\sigma,
\label{approximate propagator}
\end{flalign}
using the relation $\sigma=q/(\pi a^2)$.

Now we have the analytical solution to eq. (\ref{fourier transverse}) in the following form [25]:
\begin{flalign}
n({\bf r}_0,t)&=e^{-tD_0\Xi\,\mathcal{G}({\bf r}_0)}n({\bf r}_0,0)+\int_0^tds e^{-(t-s)D_0\Xi\,\mathcal{G}({\bf r}_0)}\zeta[\rho_{\infty}(z_0),\eta({\bf r}_0,s)].
\label{solution}%%%%%%%%%%
\end{flalign}
Considering the real space representation that $D_0\Xi\mathcal{G}({\bf r}_0)\approx 4\pi D_0\Xi\sigma$ in the above approximation of eq. (\ref{approximate propagator}), the above exponential factors, $e^{-tD_0\Xi\,\mathcal{G}}$ and $e^{-(t-s)D_0\Xi\,\mathcal{G}}\>(s<t)$, are negligible due to $\Xi\gg 1$.
Hence, eq. (\ref{solution}) is reduced to $n({\bf r}_0,t)=\zeta[\rho_{\infty}(z_0),\eta({\bf r}_0,s)]$, thereby providing
\begin{flalign}
\left<n({\bf r}_0,t)n({\bf r}'_0,t')\right>&
=\left<\zeta[\rho_{\infty}(z_0),\vec{\eta}({\bf r}_0,t)]\zeta[\rho_{\infty}({\bf r}'_0),\vec{\eta}({\bf r}'_0,t')]\right>\nonumber\\
&=2D_0\rho_{\infty}(z_0)\delta({\bf r}_0-{\bf r}'_0)\delta(t-t').
\label{white}%%%%%%%%%%%%%%
\end{flalign}
It is found from eq. (\ref{white}) that there is no correlation of transverse density fluctuations, which represents a coarse-grained frozen dynamics in the strong coupling regime of coarse-grained 2D OCP. 

We can determine a crossover scale $l_c$. or associated crossover wavenumber $k_c=2\pi/l_c$ by comparing two terms on the rhs of eq. (\ref{propagator}).
In the above approximation, the first term on the rhs of eq. (\ref{propagator}) has been neglected based on the condition that $\Xi\gg 1$ and $ka\ll 1$.
While increasing the wavenumber and maintaining the strong coupling of $\Xi\gg 1$, we arrive at the crossover wavenumber $k_c$ that is defined by the following relation:
\begin{align}
\frac{k_c^2}{\Xi}=4\pi\sigma e^{-k_c^2(ma^2)/4},
\label{crossover}
\end{align}
stating that the two terms on the rhs of eq. (\ref{propagator}) are comparable to each other. 
We now introduce the main branch $W_0(x)$ of the Lambert $W$-function \cite{lambert}, so that eq. (\ref{crossover}) is converted to
\begin{align}
\nu&=W_0(\nu e^{\nu})=W_0(qm\Xi),\nonumber\\
\nu&\equiv k_c^2(ma^2)/4,
\label{lambert1}
\end{align}
based on another expression of eq. (\ref{crossover}) as follows:
\begin{align}
\nu e^{\nu}=\pi\sigma\Xi(ma^2)=qm\Xi.
\label{crossover2}
\end{align}
The approximate form of $W_0(x)\approx\ln x$ for $x\gg 1$ \cite{lambert} applies to eq. (\ref{lambert1}) because of $qm\Xi\gg 1$.
It follows that eq. (\ref{lambert1}) reads
\begin{align}
2\pi\left(\frac{a}{l_c}\right)=
k_ca\approx
\frac{2}{m^{1/2}}\sqrt{\ln(qm\Xi)}=2.16\sqrt{\ln(qm\Xi)},
\label{lambert2}
\end{align}
which is our main result in this study.
Below this scale specified by the crossover length $l_c$, we can observe the diffusive behavior of counterions, instead of frozen correlations represented by eq. (\ref{white}).
Taking $qm\Xi=10^4$ (or $\Xi\sim 10^3$ for $q\sim 10$) as an example of the strong coupling regime, eq. (\ref{lambert2}) provides
\begin{align}
k_ca&\approx6.56,\nonumber\\
\frac{l_c}{a}&\approx\frac{2\pi}{6.56}.
\label{lambert3}
\end{align}
The latter relation implies that the transverse dynamics of strongly-coupled counterions still retain diffusive behaviors within each Wigner-Seitz cell (i.e., $l_c\sim a$), which is physically plausible (see also Fig. 3).

%%%%%%%%% FIG3 %%%%%%%%%%%%%%
\begin{figure}[hbtp]
\begin{center}
	\includegraphics[
	width=9 cm
]{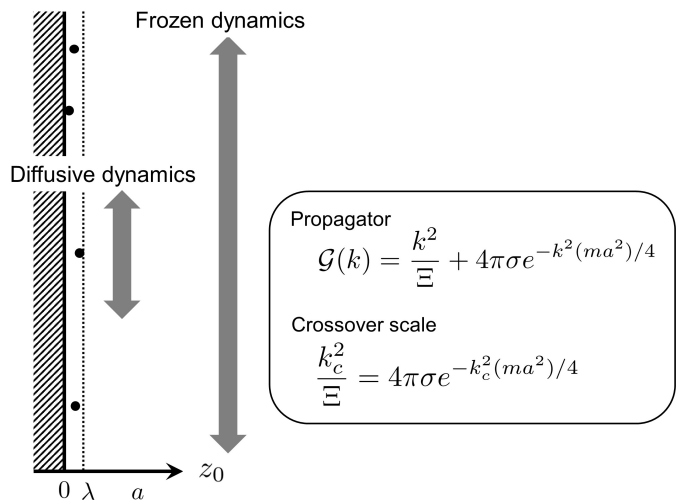}
\end{center}
\caption{
A schematic of dynamical crossover, showing that diffusive dynamics of strongly coupled counterions along the plate surface can be observed within the scale of Wigner-Seitz cell. The determining equation for the crossover scale $k_c$ is also given, based on the expression of propagator $\mathcal{G}(k)$ given by eq. (\ref{propagator}).
}
\end{figure}

%%%%%%%%%%%%%%%%%%%%%%%%%%%%%%%%%%%%%%%%%%%%%%%%%%%%%%
%%%%%%%%%%%%%%%%%%%%%%%%%%%%%%%%%%%%%%%%%%%%%%%%%%%%%%
\section{Summary and conclusions}
We have investigated stochastic density fluctuations $n=\rho-\rho_{\infty}$ around the ground state distribution ($\rho_{\infty}\propto e^{-z_0/\lambda}$) of strongly coupled counterions near a single charged plate, focusing especially on the transverse dynamics parallel to the charged plate at $z_0=0$.
The key to treating the stochastic dynamics is to use the DK equation of overdamped Brownian particles [13-27] that is linearized by expanding the first derivative of a free energy functional of given density, ($\delta F[\rho]/\delta\rho$) around the ground state density $\rho_{\infty}$.
As a result, we have obtained the linear DK equation (\ref{original stochastic}), which is applicable to the longitudinal and transverse dynamics of counterions in the strong coupling regime where the stationary density distribution has been investigated using Monte Carlo simulations [8-11].

The linear DK equation allows us to quantitatively investigate the dynamical crossover of transverse density fluctuations along the plate surface.
Accordingly, we have found a crossover scale, given by eqs. (\ref{lambert2}) and (\ref{lambert3}), above which the transverse density dynamics appear frozen, generating white noise that is uncorrelated with respect to time and space.
Below the crossover scale, on the other hand, diffusive behavior of counterions can be observed along the plate surface, as illustrated in Fig. 3.
For instance, the crossover length $l_c$ is of the order of the Wigner-Seitz radius $a$ when $\Xi\sim10^3$.
Furthermore, the longitudinal dynamics vertical to the plate arises from the gradient of a fluctuating density field along the $z$-axis, producing additional contributions to the transverse dynamics, such as electrically reverse flow as well as advective flow (see Fig. 2).
The electrically reverse flow would be crucial in experimental situations where mobile ions, including not only counterions but also added salt, are affected considerably by the longitudinal dynamics. 
This remains to be addressed in a quantitative manner, by extending the present formulation to multi-component systems.

\appendix
%%%%%%%%%%%%%%%%%%%%%%%%%
%%%%%%%%%%%%%%%%%%%%%%%%%
\section{Electrostatic interaction energies: general forms when rescaled by the Wigner-Seitz radius $a$}
 The counterion number $N$ is imposed on the global electroneutrality condition, $-qN+\sigma\Sigma=0$ ($\Sigma$: the surface area of a macroion). The system has electrostatic interaction energy $U=U_{cc}+U_{cm}+U_{mm}$, i.e., the sum of the counterion-counterion interaction energy $U_{cc}$, the counterion-macroion interaction energy $U_{cm}$, and the self-energy energy of macroion $U_{mm}$ due to electrostatic interactions between charged groups fixed on the macroion surface. In the thermal energy unit, we have
    \begin{eqnarray}
    U_{cc}\{{\bf r}_{01},\cdots,{\bf r}_{0N}\}
    &=&\frac{q^2\,l_B}{2}\left(
    \sum_{i,j=1}^N\,v({\bf r}_{0i}-{\bf r}_{0j})-N\,v(0)
    \right)\nonumber\\
    U_{cm}\{{\bf r}_{01},\cdots,{\bf r}_{0N}\}
    &=&-q\,l_B\,\sigma\sum_{i=1}^N\,\oint\,dS_0\>v({\bf r}_{0i}-{\bf R}_0),
    \label{energetics1}
    \end{eqnarray}
where the interaction potential $v({\bf r}_0)$ in the $k_BT$ unit is represented by an ion-ion separation vector, ${\bf r}_0=(x_0,y_0,z_0)$, as $v({\bf r}_0)\equiv1/r_0$ with $r_0=|{\bf r}_0|$ , and $\oint\,dS$ denotes that the integration of the macroion charge position ${\bf R}_0$ is restricted to the surface of the macroion.

Let us rescale the system as ${\bf r}=(x,y,z)=(x_0/a,y_0/a,z_0/a)={\bf r}_0/a$, using the mean counterion-counterion separation $a$ (or the Wigner-Seitz radius $a$) that is evaluated from the electrical neutrality condition, $\pi a^2\sigma=q$, supposing that all counterions located on the macroion surface are uniformly distributed.
Since the coupling constant $\Gamma$ defined by eq. (\ref{gamma}) gives $ql_B\sigma=q^2l_B/(\pi a^2)=\Gamma/(\pi a)$, eq. (\ref{energetics1}) reads
\begin{eqnarray}
U\{{\bf r}_1,\cdots,{\bf r}_N\}&=&U_{cc}\{{\bf r}_1,\cdots,{\bf r}_N\}+U_{cm}\{{\bf r}_1,\cdots,{\bf r}_N\}+U_{mm}\nonumber\\
    U_{cc}\{{\bf r}_1,\cdots,{\bf r}_N\}
    &=&\frac{\Gamma}{2}\left(
    \sum_{i,j=1}^N\,v({\bf r}_i-{\bf r}_j)
    -Nv(0)
    \right)
    \nonumber\\
    U_{cm}\{{\bf r}_1,\cdots,{\bf r}_N\}
    &=&-\frac{\Gamma}{\pi}\sum_{i=1}^N\,\oint\,dS\>
    v({\bf r}_i-{\bf R}),
    \label{energetics3}
    \end{eqnarray}
where we have used the relation $dS=dS_0/a^2$.
Equation (\ref{energetics3}) implies that the electrostatic interaction energies become extremely large in the strong coupling limit of $\Gamma\rightarrow\infty$.

The 2D OCP formed on the macroion surface has been often regarded as a ground state of strongly coupled counterions.
For incorporating this analogy to the ground state of the 2D OCP into our formulation, we set the base energy $U\{{\bf R}_1,\cdots,{\bf R}_N\}$ given by
  \begin{align}
  U\{{\bf R}_1,\cdots,{\bf R}_N\}&=U_{cc}\{{\bf R}_1,\cdots,{\bf R}_N\}+U_{cm}\{{\bf R}_1,\cdots,{\bf R}_N\}+U_{mm}.
\label{base}%%%%%%%%%%%%%%%%%
  \end{align}
We write $U_{mm}$, a constant interaction energy, as $U_{mm}=Nu_m$ using the interaction energy $u_{mm}$ of one charged group located at a reference position of ${\bf R}_{\mathrm{ref}}$, which is expressed as
   \begin{equation}
   u_{mm}=\frac{\Gamma}{2\pi}\oint\,dS\>
    v({\bf R}_{\mathrm{ref}}-{\bf R}).
    \label{rescale-energy}
   \end{equation}
The base energy vanishes ($ U\{{\bf R}_1,\cdots,{\bf R}_N\}\approx 0$) in the approximation that
  \begin{align}
  U_{cc}\{{\bf R}_1,\cdots,{\bf R}_N\}&\approx Nu_{mm}\nonumber\\
  U_{cm}\{{\bf R}_1,\cdots,{\bf R}_N\}&\approx -2Nu_{mm}.
\label{counterion base}
  \end{align}
It is also convenient to introduce the reference energy $\overline{U}_m$ associated with the charged groups on macrion surface:
  \begin{align}
U\{{\bf R}_1,\cdots,{\bf R}_N\}&=U_{cc}\{{\bf R}_1,\cdots,{\bf R}_N\}+ \overline{U}_m\nonumber\\
\overline{U}_m&\equiv U_{cm}\{{\bf R}_1,\cdots,{\bf R}_N\}+U_{mm}\approx -Nu_{mm},
  \end{align}
where the reference energy $\overline{U}_m$ can be regarded as a constant energy, as found from the above approximation in eq. (\ref{counterion base}).

%%%%%%%%%%%%%%%%%%%%%%%%%
%%%%%%%%%%%%%%%%%%%%%%%%%
\section{The grand potential $\Omega[J]$ for a one-plate system}
To clarify the underlying physics behind the counterion-macroion electrostatic interactions in a one charged plate system, it is useful to consider the difference between the actual energy $U_{cm}\{{\bf r}_1,\cdots,{\bf r}_N\}$ and the base energy $U_{cm}\{{\bf R}_1,\cdots,{\bf R}_N\}$ in a condensed state that all counterions are attached to the plate with their locations distributed uniformly:
 \begin{align}
\Delta U_{cm}\{{\bf r}_1,\cdots,{\bf r}_N\}
&=U_{cm}\{{\bf r}_1,\cdots,{\bf r}_N\}-U_{cm}\{{\bf R}_1,\cdots,{\bf R}_N\}\nonumber\\
&=\sum_{i=1}^N J({\bf r}_i),
\label{delta_ucm}%%%%%%%%%%%%%
\end{align}
where $J({\bf r}_i)$ denotes the external potential that is experienced by the $i$-th counterion due to the one charged plate and is defined by
\begin{equation}
J({\bf r}_i)=\frac{\Gamma}{\pi}\oint\,dS\>
    \left\{
v({\bf r}_i-{\bf R})-v({\bf R}_i-{\bf R})
\right\}.
\label{macroion_potential}%%%%%%%%%%%%%%%%
\end{equation}
In the coordinate setting of a schematic system depicted in Fig. 1, we have ${\bf R}={\bf r}\delta(z)$ and $J({\bf r})$ simply reads
\begin{equation}
J({\bf r})=\frac{z_0}{\lambda}=z\left(\frac{a}{\lambda}\right)
=2\Gamma z,
\label{one-plate}%%%%%%%%%%%%%%%%%%
\end{equation}
where we have used the relation (\ref{length_comparison}) in the above last equality.

The electrostatic interaction energy $U$ given by eq. (\ref{energetics3}) thus reads, for a one-plate system,
\begin{equation}
U\{{\bf r}_1,\cdots,{\bf r}_N\}=U_{cc}\{{\bf r}_1,\cdots,{\bf r}_N\}
+\overline{U}_m
+\Delta U_{cm}\{{\bf r}_1,\cdots,{\bf r}_N\}.
\label{appendix energy one plate}
\end{equation}
We can now define the grand potential $\Omega[J]$ of the counterion system under the external field of $J({\bf r})$ created by the one charged plate.
The configurational representation of $\Omega[J]$ is represented as
%%%%%%%%%%%%%%%%%%%%%%(2)
\begin{align}
&e^{-\Omega[J]}=e^{-\overline{U}_m}\mathrm{Tr}\,\exp\left\{
-U_{cc}\{{\bf r}_1,\cdots,{\bf r}_N\}
-\sum_{i=1}^NJ({\bf r}_i)
\right\}\nonumber\\
&\mathrm{Tr}\equiv
\sum_{N=0}^{\infty}e^{N\beta\mu}\frac{1}{N!}\int d{\bf r}\,_1\cdots\int d{\bf r}\,_N,
\label{f-start}%%%%%%%%%%%%%%%%%%%%
\end{align}
where $\mu$ denotes chemical potential and the expression (\ref{delta_ucm}) has been used.

%%%%%%%%%%%%%%%%%%%%%%%%%
%%%%%%%%%%%%%%%%%%%%%%%%%
\section{A remark on eq. (\ref{approximate f rho}) \cite{frusawa2019}}

Going beyond the mean-field approximation, the free energy functional $F[\rho]$ is not identified with $\mathcal{A}[\rho]$.
We have an additional contribution $\Delta F[\rho]$ to $F[\rho]$ that is obtained from the functional integration over fluctuating potential field $\phi=\psi-\psi^*$ around the saddle-point field $\psi^*$ determined by eq. (\ref{sp}):
%%%%%%%%%%%%%%%%%%%%%%(6)
\begin{flalign}
&e^{-\Delta F[\rho]}=\int D\phi\,e^{-\Delta\mathcal{H}[\rho,\phi]},
\label{hk}%%%%%%%%%%%%%%%%%
\end{flalign}
where
\begin{align}
\Delta\mathcal{H}[\rho,\phi]&=\Delta\Omega+\int d{\bf r}\,i\rho({\bf r})\phi({\bf r}),\nonumber\\
\Delta\Omega&=\Omega[-i\phi+\psi^*]-\Omega[\psi^*].
\label{additional omega}
\end{align}
The free energy functional $F[\rho]$ of a given density is thus written as
\begin{equation}
F[\rho]=\mathcal{A}[\rho]+\int d{\bf r}J({\bf r})\rho({\bf r})+\Delta F[\rho].
\label{given density}
\end{equation}
In the Gaussian approximation, $\Delta F[\rho]$ corresponds to the logarithmic correction term \cite{frusawa2019}, which we have neglected in eq. (\ref{approximate f rho}) as the first approximation of the strong coupling regime of $\Gamma\gg 1$.

%%%%%%%%%%%%%%%%%%%%%%%%%%%%%%%%%%%%%%%%%%%%%%%%%%%%%%%%%%%%%%%%%%%%%%%%%%%%%%%%%%%%%%%%%%%%%%%%%%%%%%%%%%%%%%%%%%%
\section{Details of eq. (\ref{constant first derivative})}
In the mean-field approximations of eqs. (11) to (13), we have
\begin{flalign}
\left.
\frac{\delta F[\rho]}{\delta\rho}\right|_{\rho=\rho_{\infty}}
&=\left.
\frac{\delta\mathcal{A}[\rho]}{\delta\rho}\right|_{\rho=\rho_{\infty}}
+J({\bf r}_0)\nonumber\\
&=\ln\rho_{\infty}(z_0)-c^{(1)}({\bf r}_0,\rho_{\infty})-u_{mm}+J({\bf r}_0)\nonumber\\
&=\ln\rho_{\infty}(0)-c^{(1)}({\bf r}_0,\rho_{\infty})-u_{mm}\nonumber\\
&\approx \ln\rho_{\infty}(0),
\label{first derivative}
\end{flalign}
where the first member of the hierarchy of the direct correlation function, $c^{(1)}({\bf r},\rho_{\infty})$, corresponds to the effective potential due to counterion-counterion interactions, and the last approximate equality ignores the difference $-c^{(1)}({\bf r},\rho_{\infty})-u_{mm}$ between the effective potentials created by condensed counterions and charged plate, in comparison with $\ln\rho_{\infty}(0)=\ln(\sigma/\lambda)$.
Equation (\ref{first derivative}) indicates that spatial dependence of $\delta F[\rho]/\delta\rho|_{\rho=\rho_\infty}$ is negligible, thereby verifying eq. (\ref{constant first derivative}).
\end{document}